# Highly Tunable Perpendicular Magnetic Anisotropy and Anisotropic Magnetoresistance in Ru-doped La$_{0.67}$Sr$_{0.33}$MnO$_3$ Epitaxial Films

Enda Hua[1], Kunjie Dai[1], Qing Wang[1], Huan Ye[1], Kuan Liu[1], Jinfeng Zhang[1], Jingdi Lu[1], Kai Liu[1], Feng Jin[1], Lingfei Wang[1],* and Wenbin Wu[1,2] *

[1]Hefei National Research Center for Physical Sciences at Microscale, University of Science and Technology of China, Hefei 230026, China

[2]Collaborative Innovation Center of Advanced Microstructures, Nanjing University, Nanjing 210093, China

E-mail: wanglf@ustc.edu.cn, wuwb@ustc.edu.cn

**Abstract**

As a prototypical half-metallic ferromagnet, La$_{0.67}$Sr$_{0.33}$MnO$_3$ (LSMO) has been extensively studied due to its versatile physical properties and great potential in spintronic applications. However, the weak perpendicular magnetic anisotropy (PMA) limits the controllability and detection of magnetism in LSMO, thus hindering the realization of oxide-based spintronic devices with low energy consumption and high integration level. Motivated by this challenge, we develop an experimental approach to enhance the PMA of LSMO epitaxial films. By cooperatively introducing 4*d* Ru doping and a moderate compressive strain, the maximum uniaxial magnetic anisotropy in Ru-doped LSMO can reach $3.0 \times 10^5$ J/m$^3$ at 10 K. Furthermore, we find a significant anisotropic magnetoresistance effect in these Ru-doped LSMO films, which is dominated by the strong PMA. Our findings offer an effective pathway to harness and detect the orientations of magnetic moments in LSMO films, thus promoting the feasibility of oxide-based spintronic devices, such as spin valves and magnetic tunnel junctions.

## 1. Introduction

Magnetic anisotropy (MA), referring to the "easy" and "hard" magnetization directions, is one of the most important magnetic properties in ferromagnets.[1–6] With the rapid development of spintronics, investigating MA becomes more and more critical for fabricating highperformance spintronic devices, such as magnetic tunnel junctions, non-volatile magnetic memories, and magnetic sensors.[7,8] In ferromagnetic thin films and devices, perpendicular magnetic anisotropy (PMA), defined by the magnetic easy axis along the film normal, is particularly vital for high-density storage and high-sensitivity magnetic field detection.[9–13] Moreover, samples with PMA can also enable precise characterizations of the anomalous Hall effect (AHE), thereby facilitating effective readout of magnetic information.[14] Owing to these appealing advantages, researchers have devoted intensive efforts to inducing and enhancing the PMA in magnetic thin films and heterostructures. Typical examples include strain engineering, interface engineering, and inserting heavy metal layers with strong spin-orbit coupling (SOC).[9,15–20]

Optimal doped manganite $La_{0.67}Sr_{0.33}MnO_3$ (LSMO) is a typical transition-metal oxide with robust room-temperature metallicity and ferromagnetism. The Curie temperature $T_C$ and saturated magnetization $M_s$ can be as high as 370 K and 3.7 $\mu_B$/Mn, respectively. As a typical strong electron–electron correlation system, LSMO exhibits a wide range of fascinating properties, including colossal magnetoresistance (CMR), half-metallicity, and electronic phase separation.[21–24] Moreover, the strong coupling between spin, charge, orbital, and lattice degrees of freedom further makes these physical properties highly tunable under external stimuli. Therefore, LSMO has been widely used as electrode layers for oxide-based heterostructures. Nevertheless, LSMO has small coercive field $H_c$ and weak magnetocrystalline anisotropy, which considerably restricts its device applications. For instance, in LSMO-based magnetic tunnel junctions, the small magnetic moment in the pining layer is unstable under an external magnetic field, and the in-plane magnetic moment caused by the demagnetization effect limits the device size. To overcome these limitations, researchers have developed various experimental approaches to induce and enhance the PMA in LSMO.[3,17,25] One of the most straightforward approaches is using compressive strain. By growing LSMO on $LaAlO_3$(001) substrate with a smaller lattice constant, over 2% compressive misfit strain can induce additional Jahn–Teller distortion to the $MnO_6$ octahedron. This distortion results in a preferential occupancy of the $3z^2–r^2$ orbital[26–29] and thus induces PMA through SOC.[26] However, the large compressive strain also hinders the double-exchange interactions and weakens the ferromagnetism. The $M_s$ of LSMO/LaAlO$_3$(001) epitaxial films degrades from the bulk value of ~ 3.7 $\mu_B$/Mn to ~ 1.4 $\mu_B$/Mn at 5 K.[26] Therefore, developing an effective route to simultaneously enhance the PMA and to maintain the strong ferromagnetism of LSMO epitaxial films is highly required.

In this study, we introduce a simple and effective approach for enhancing the PMA of LSMO epitaxial films: doping with 4$d$ element Ru. In the La$_{0.67}$Sr$_{0.33}$(Mn$_{1-x}$Ru$_x$)O$_3$ (LSMRO) films grown on (001)-oriented (LaAlO$_3$)$_{0.3}$(SrAl$_{0.5}$Ta$_{0.5}$O$_3$)$_{0.7}$ [LSAT(001)] substrates, the Ru dopants significantly amplify the SOC and thus magnify the compressive strain effect on the PMA. The maximum uniaxial magnetocrystalline anisotropy constant $K_u$ at 10 K reaches up to $3.0 \times 10^5$ J/m$^3$, surpassing most of the LSMO films reported previously. Moreover, we also observed a significant anisotropic magnetoresistance (AMR) effect in these LSMRO/LSAT(001) films. Namely, the sign and magnitude of magnetoresistance (MR) are sensitive to the direction of the external magnetic field. Such an AMR effect is strongly correlated with the PMA in the LSMRO/LSAT(001) films. Our findings suggest that the Ru doping serves as an effective strategy to enhance both the PMA and AMR of LSMO epitaxial films, opening a new avenue for promoting the potential of LSMO for spintronic device applications.

## 2. Method and Material

We employed pulsed laser deposition (PLD) to grow various Ru-doped LSMO ($0 \leq x \leq 0.20$) films on LSAT(001) substrates. The LSMRO ceramic target was obtained by sintering high-purity powders in an appropriate stoichiometric ratio at 1350 °C for 24 h using a solid-state sintering method. During the sample growth, we maintained the substrate temperature at 680 °C, and the oxygen atmosphere at 45 Pa. The laser fluence used for ablating the target was about 2 J·cm$^{-2}$. After deposition, we first annealed the film *in situ* under the growth condition for 15 min and then cooled the samples down to room temperature in an oxygen environment.

We performed high-resolution X-ray diffraction (XRD, PANalytical Empyrean X-ray diffractometer, Cu $K_{\alpha 1}$ radiation) to characterize the high-quality epitaxial structure of the samples. Characterizations of magnetization were carried out using a vibrating sample magnetometer (VSM-SQUID, Quantum Design), and characterizations of the electrical transport properties were conducted using a physical properties measurement system (PPMS, Quantum Design). To measure the longitudinal and Hall resistance, we patterned the film samples into Hall bars using photolithography.

## 3. Results and Discussion

We first characterized the epitaxial quality and strain state of the LSMRO epitaxial films. The XRD $2\theta$–$\omega$ linear scans (Fig. 1a) of LSMRO/LSAT(001) films show strong LSMRO(002) diffractions and clear Laue fringes, indicating a sharp interface between the substrate and the film. We find that the rhombohedral symmetry of LSRMO remains unchanged across the entire Ru doping range, while the LSAT(001) substrate has a cubic structural symmetry. As Ru doping increases, the position of the LSMRO(002) diffractions shifts gradually toward smaller Bragg angles, indicating lattice expansion along the out-of-plane direction. We further characterized the film structure through offspecular reciprocal space mappings (RSMs). As shown in Fig. 1b–d, the in-plane reciprocal space vectors $Q_x$ for all of the LSMRO films are the same as those of LSAT(001) substrates, demonstrating coherent strain states. The observed gradual reduction of out-of-plane reciprocal space vector $Q_z$ in LSMRO films further suggests that the out-of-plane lattice constant of the LSMRO film increases with Ru doping, consistent with the trend indicated by XRD $2\theta$–$\omega$ linear scans. Such a lattice expansion is likely due to the larger ionic radius of Ru cations compared to Mn cations.[30] Additionally, we plot the lattice mismatch $\varepsilon_f$ and out-of-plane $d$-spacing of the LSMRO(002) plane [$d_{LSMRO(002)}$] as functions of the Ru doping level $x$, further revealing a gradually increased compressive strain with $x$ (Fig. 1e).

The MA of LSMRO/LSAT(001) films is highly dependent on the Ru doping level $x$. We performed detailed magnetic characterization of these LSMRO samples with the external magnetic field $H$ applied both inplane ($H_\parallel$) and along the film normal ($H_\perp$) direction, respectively. As shown in Fig. 2a–2c, all of the $M$–$T$ curves exhibit bifurcation behaviors. For the LSMRO film with $x = 0.10$ and 0.15, at low temperature, the $M$ values for the $H_\perp$ cases are always higher than that in the $H_\parallel$ cases, suggesting a preferential alignment of spins along the film normal, namely, a PMA. The $M$–$T$ curves measured from the LSMO film show an opposite trend, signifying a magnetic easy plane anisotropy. Moreover, the bifurcation behaviors always disappear as the temperature increases above a certain value, which suggests that the MA strongly depends on the temperature. The evolution of MA can be further characterized by the field-dependent magnetization hysteresis ($M$–$H$) at 10 K (Fig. 2d–2f). For the LSMO/LSAT(001) film, as $H_\parallel$ increases, $M$ saturates faster in the $H_\parallel$ case. This result further confirms the magnetic easy plane anisotropy, probably dominated by the demagnetization effect. For the LSMRO/LSAT(001) film with $x = 0.10$, $M$ saturates slightly faster in the $H_\perp$ case, and the $H_c$ becomes larger ompared to the parent LSMO/LSAT(001) film, signifying a moderate PMA. For the LSMRO/LSAT(001) film with $x = 0.15$, the $M-H_\parallel$ hysteresis loop saturates at a higher $H$ value, while the $M-H_\perp$ hysteresis loop exhibits a square shape with larger $H_c$, signifying robust PMA, in contrast to the LSMO/LSAT(001) film.

To further investigate the impact of Ru doping on PMA, we calculated the uniaxial anisotropy constant ($K_u$) using the formula:[31,32]

$$K_u = K_{eff} + 2\pi M_s^2$$

where $K_{eff}$ is the effective anisotropy constant; $2\pi M_s^2$ is the demagnetization energy. $K_{eff}$ can be estimated as $K_{eff} = M_s \cdot H_a/2$, with $M_s$ and $H_a$ representing the saturated magnetization and anisotropic field, respectively. The positive (negative) sign $H_a$ of films is defined as the magnetic easy axis align out-of-plane (in-plane), and the value of $H_a$ is calculated by the difference in saturation fields ($H_s$) between the $H_\parallel$ and $H_\perp$ cases. As shown in Fig. 2g, $K_{eff}$ increases with $x$ and becomes positive as $x = 0.05$, indicating a clear Ru-doping-induced transition from the magnetic easy plane anisotropy to the PMA. Additionally, Fig. 2g (blue curve) displays the calculated value of $K_u$ as a function of $x$. The results show that the $K_u$ values for all the Ru doping levels are positive, which can be attributed to the compressive strain imposed by the LSAT(001) substrate. Notably, 5% Ru doping can trigger a sudden rise of $K_u$ and then it increases gradually with $x$. The maximum $K_u$ of $3 \times 10^5$ J/m$^3$ can be achieved in the LSMRO ($x = 0.15$) film at 10 K. To the best of our knowledge, such a strong PMA has not been realized in LSMO epitaxial films so far.[33]

The epitaxial strain should play a dominating role in inducing PMA in manganite epitaxial films. According to previous reports, the $K_u$ value in the LSMO/LAO(001) film can reach ~ $1 \times 10^5$ J/m$^3$, induced by a huge compressive strain over 2%.[26,33] Such a large compressive strain in LSMO epitaxial films can induce Jahn–Teller distortion of the MnO$_6$ oxygen octahedra and facilitate a preferential $d_{3z^2-r^2}$ orbital occupancy of the $e_g$ electrons, thereby resulting in PMA through SOC.[25] For the LSMRO film with $x = 0.15$, although displaying a strong PMA, the compressive strain imposed by LSAT(001) is only 0.465%. On this basis, we suggest that Ru doping can effectively "amplify" the strain effect on magnetocrystalline anisotropy and thus the PMA. Namely, the 4$d$ Ru dopants could effectively enhance the SOC strength and thus bolsters the alignment of spins along the orbital polarization direction.[34] This scenario can explain why even a modest amount of compressive strain (less than 0.5%) can trigger a strong PMA in LSMRO/LSAT(001) films. In brief, the enhancement of PMA in LSMRO films is a collective effect of compressive strain and Ru doping.

To further understand the highly tunable MA in LSMRO, we conducted angle-dependent magnetotransport measurements. As schematically depicted in Fig. 3a, during the resistivity ($\rho$) measurements, we rotated the $H$ from the film normal to the in-plane direction and kept the current direction always in the $H$ rotation plane. We defined the angle between the film normal and $H$ as $\theta$. For LSMRO samples with different $x$, AMR was characterized under a constant $\mu_0 H = 3$ T, which is strong enough to saturate the magnetic moment along the $H$ direction. Fig. 3b–d display the polar plots of AMR for LSMO and LSMRO samples. AMR was calculated using the formula: AMR = [($\rho_\theta - \rho_{min}$)/ $\rho_{min}$] ×100%, where $\rho_\theta$ and $\rho_{min}$ are resistivities measured at $\theta$ and the minimum value, respectively. All of the AMR polar plots exhibit two-fold symmetry, whereas the plots of LSMO and LSMRO samples show a clear 90° phase difference. The LSMO/LSAT(001) film shows a higher AMR at $\theta = 0°$ ($H_\perp$) than that at $\theta = 90°$ ($H_\parallel$). In contrast, LSMRO samples display maximum AMR at $\theta = 90°$ and minimum AMR at $\theta = 0°$. As mentioned above, the magnetic easy axis of the LSMO film lies in the film plane ($\theta = 90°$), while it rotates to the film normal ($\theta = 0°$) for the LSMRO ($x = 0.10, 0.15$) films. Namely, the AMR results are consistent with the observed switching of the magnetic easy axis. In addition to the phase shift, notably, the magnitude of the AMR also changes significantly with Ru doping level $x$. As shown in Fig. 3b, the AMR of LSMO is negligible at 10 K and increases gradually to ~ 1% at 300 K. For the LSMRO ($x = 0.10$) film, the AMR at 10 K increases to ~ 10%. For the LSMRO film with $x = 0.15$, AMR increases further to a significant value up to ~ 40%. To clarify the evolution of AMR, we characterized the MR at different $H$ directions (i.e., different $\theta$ values). The MR is calculated using the equation MR = [$\rho(H) - \rho(H = 0)$]/ $\rho(H = 0)$. Fig 3e–g display the MR–$H$ curves of LSMO and LSMRO films measured with various $\theta$ values at 10 K. The MR curves can be detangled into two components: (1) the hysteric MR at low field (LFMR), corresponding to the magnetic switching behavior, and (2) the non-hysteric MR observed in the high field region (HFMR). As shown in Fig. 3e, the LSMO/LSAT(001) film shows negative MR for all the $\theta$ values. The HFMR exhibits a nearly linear $H$-dependence, and the maximum value is only −0.3% at ±2.5 T for the $\theta = 90°$ case. The MR also decreases lightly as $\theta$ rotates to 0°. For the LSMRO films, consistent with the 90° rotation of the magnetic easy axis, the MR also displays an opposite trend. For the film with $x = 0.10$, the HFMR exhibits negative MR at $\theta = 0°$. As $\theta$ increases, HFMR gradually transforms to positive, reaching the maximum value up to ~ 8% at $\theta = 90°$. The MR–$H$ curve measured at $\theta = 90°$ shows a notable deviation from the linear trend in the high $H$ region. The LSMRO film with $x = 0.15$ also exhibits a similar sign reversal from negative to positive as $\theta$ increases from 0 to 90°, but the HFMR becomes more linear. The maximum value can reach ~ 25% at ±2.5 T. The evolution of HFMR with $x$ is consistent with the AMR results.

In the previous work, we have discovered that Ru doping can enhance the AHE in LSMRO.[35] Based on this consideration, we measured the $H$-dependent anomalous Hall resistivity ($\rho_{AHE}$–$H$) at different $H$ directions. As shown in Fig. 4a,b, the $\rho_{AHE}$–$H$ curves can perfectly merge into one curve by converting the $H$ to the out-of-plane component $H\cos\theta$. This observation implies that the AHE magnitude depends solely on the out-ofplane component of $H$. However, the MR curves measured at different $H$ directions show distinct line shape changes and even sign reversal, suggesting a more complex origin. Namely, the evolutions of AMR are highly correlated to the MA. Following this scenario, we measured the MR–$H$ curves of LSMRO ($x = 0.15$) films at various temperatures. As shown in Fig. 4c–e, the MR–$H$ curves show two obvious evolutions as temperature increases from 10 to 300 K. First, the HFMR measured at large $\theta$ values (near films plane) evolves from positive to negative. Second, the MR–$H$ curves evolve from anisotropic to isotropic. The isotropic and negative MR measured at 300 K is consistent with the negligible PMA shown in the $M$–$H$ curves (Fig. 4f). These results further demonstrate the strong correlation between PMA and AMR in LSMRO films.

We now discuss the origin of AMR in LSMRO films. As aforementioned, the MA should play a critical role in inducing the AMR. For ferromagnetic manganites, the conduction mechanisms and local magnetic mo ments are strongly coupled through the double-exchange interaction.[36,37] In general, the MR should mainly originate from the scattering of itinerant electrons by the local magnetic moments. The spin-polarization direction of itinerant electrons should always follow the direction of external $H$, while the local magnetic moments align with the magnetic easy axis when $H$ is small, and tend to be parallel to the external $H$ as the $H$ strength increases. The LSMO/LSAT(001) film has a weak in-plane MA. As the $H_\parallel$ ($\theta = 90°$) increases, the spin-polarized itinerant electrons tend to align with the local magnetic moment. The alignment becomes better as $H_\parallel$ increases, thus leading to the negative MR. For the $\theta = 0°$ case, as the $H_\perp$ increases, it can easily overcome the demagnetization effect and align the local moment and spin-polarized electrons, resulting in a similar negative MR. For the LSMRO films with strong PMA, the scattering processes of itinerant electrons should be quite different. As the $H_\perp$ ($\theta$ = 0°, along the magnetic easy axis) increases, the alignment of local magnetic moments and spin-polarization of itinerant electrons also cause a negative MR. By contrast, as the applied $H$ rotates towards the in-plane direction and gradually deviates from the magnetic easy axis, the spin polarization direction of itinerant electrons (which always follow the $H$ direction) progressively misaligned with the local magnetic moments. This misalignment inevitably increases the scattering rate of itinerant electrons.[38–40] When $H$ is inadequate to overcome the strong PMA, the scattering should increase with $H$, thus leading to the positive MR.[41,42] As the Ru doping level $x$ increases, the enhanced PMA is expected to enlarge the positive MR at higher $H$. Also, as temperature increases, the degraded PMA is expected to reduce the positive MR. A high $H$ can overcome the weak PMA and convert the MR sign from positive to negative. Moreover, the Ru doping can also enhance the spin frustration in LSMRO films, which may further enhance the spin-dependent scattering and thus the positive MR.[35,42]

4. **Summary**

In summary, we have developed a feasible experimental approach to enhance the PMA in LSMO films through Ru doping. The 4$d$ Ru dopants can enhance the SOC in LSMO and amplify the orbital polarization-denominated magnetocrystalline anisotropy. The maximum $K_u$ can reach 3.0 × 10$^5$ J/m$^3$ at 10 K in the LSMRO film with $x$ = 0.15, which has not been achieved in ferromagnetic manganites by epitaxial strain only. The resultant PMA dramatically changes the scattering process of itinerant electrons in LSMRO films, leading to a field directionsensitive AMR effect. The robust PMA and the unique AMR effect together provide a solid foundation for potential applications in spintronic devices.

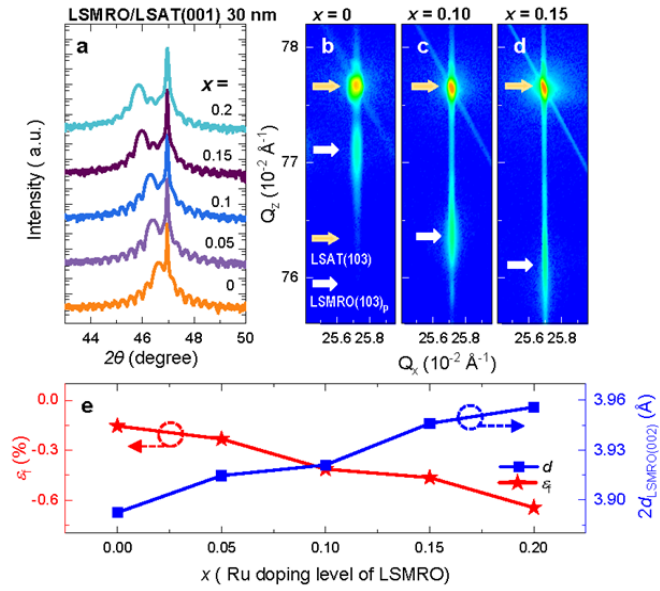

**Figure 1. Epitaxial quality and strain state of LSMRO epitaxial films.** a) The XRD *2θ-ω* linear scans near LSMRO(002) diffractions, measured from 30 nm LSMRO/LSAT(001) films with different Ru doping level *x*. b-d) Off-specular reciprocal space mappings (RSMs) near LSAT(103) diffractions. e) Lattice mismatch $\varepsilon_f$ and out-of-plane *d*-spacing $d_{\text{LSMRO}(002)}$ of LSMRO films plotted as functions of Ru doping level *x*.

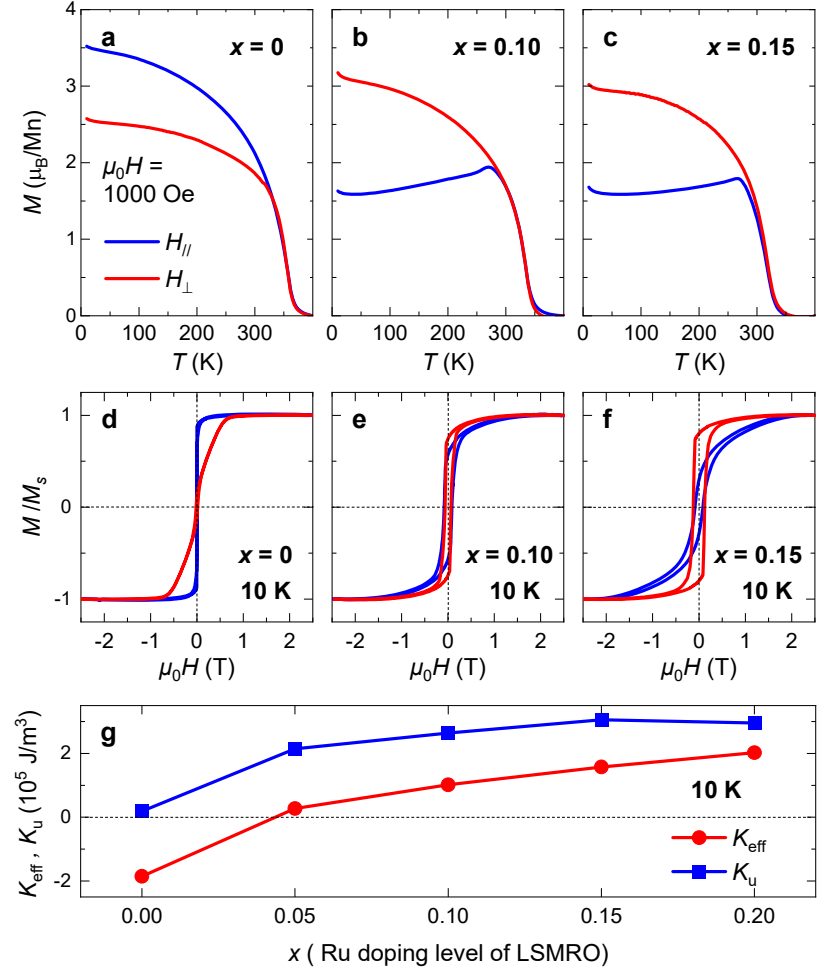

**Figure 2. Characterization and analysis of magnetic anisotropy in LSMRO/LSAT(001) films.** a-c) The temperature-dependent magnetization (*M-T*) curves of LSMRO/LSAT(001) films with various Ru doping level *x*. During the measurements, a static magnetic field $\mu_0 H = 1000$ Oe is applied in-plane ($H_{//}$, blue) or along the film normal ($H_\perp$, red). d-f) *H*-dependent magnetization (*M-H*) curves measured from LSMRO films at 10 K. g) Calculated effective magnetic anisotropy constant $K_{\text{eff}}$ and uniaxial magnetic anisotropy constant $K_{\text{u}}$ with various Ru doping level *x*.

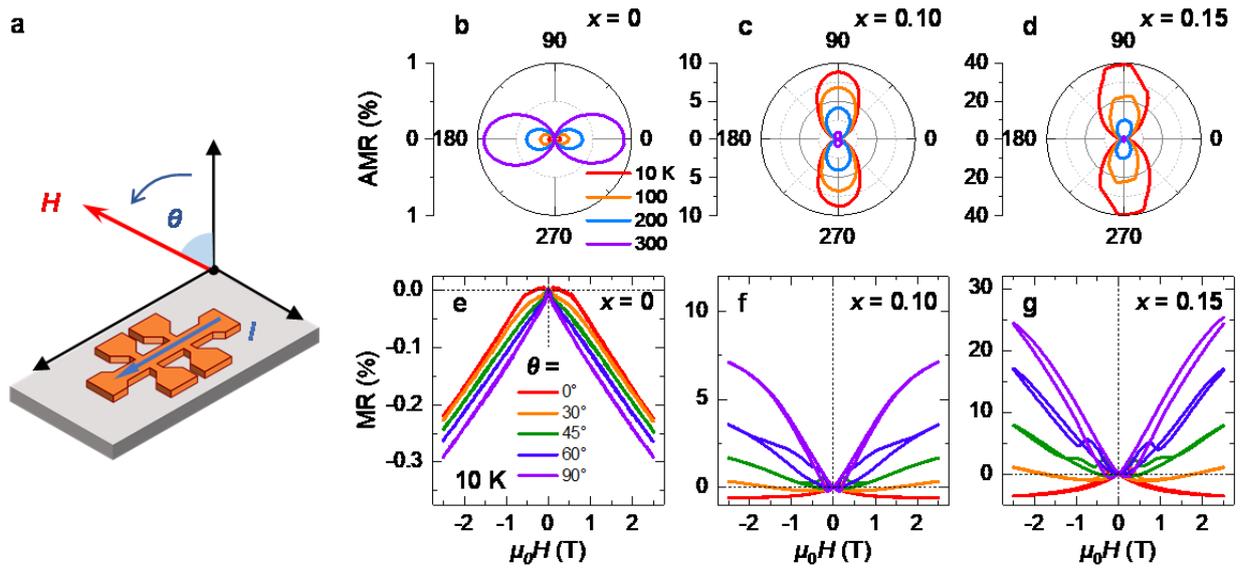

**Figure 3. Anisotropic magnetoresistance of LSMRO/LSAT(001) films.** a) Schematic experiment setup for AMR measurement. b-d) Polar plots of AMR at different temperatures, measured from LSMRO films with various $x$. e-g) $H$-dependent magnetoresistance curves (MR-$H$) at different $\theta$ values ($H$ directions) for LSMRO films with various $x$.

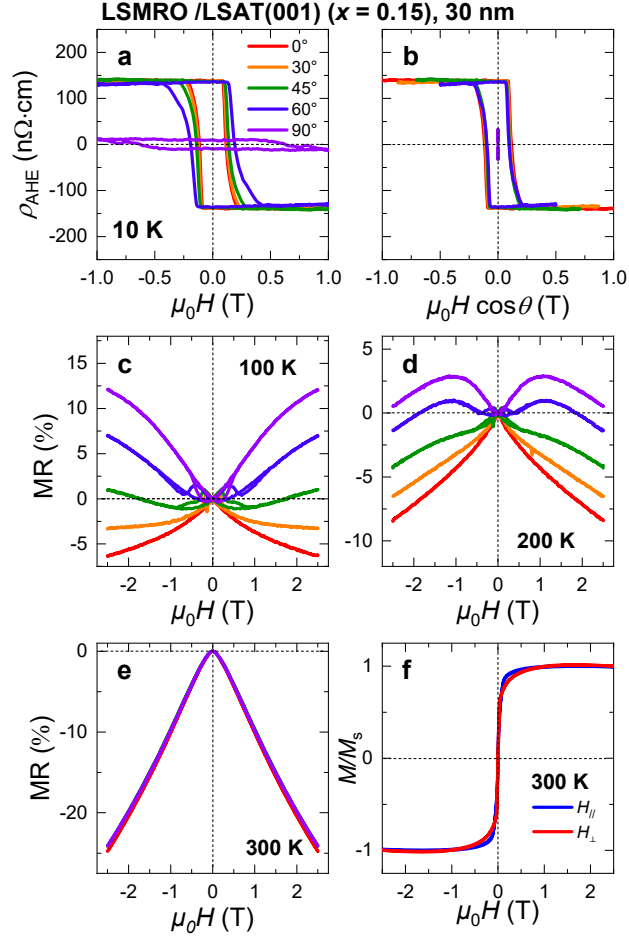

**Figure 4. Correlation between AMR and MA in LSMRO films.** a) $H$-dependent anomalous Hall resistivity ($\rho_{AHE}$-$H$) curves measured at 10 K on LSMRO films ($x$ = 0.15) with different $H$ directions. b) $\rho_{AHE}$-$H\cos\theta$ with the out-of-plane component of $H$ based on the curves of Fig. 4a. c-e) The curves of $H$-dependent MR (MR-$H$) with different $H$ directions at various temperatures. f) The $H$-dependent magnetization ($M$-$H$) curves measured at 300 K.